\documentclass[twoside]{dis08}
\usepackage[latin1]{inputenc}
\usepackage[dvips]{graphicx,epsfig,color}
\usepackage{wrapfig,rotating}
\usepackage{amssymb,amsmath,array}
\usepackage{cite}

\begin{document}
\title{High Precision Gauge Boson Pair Production at the LHC}

\author{G. Chachamis
%
%
\vspace{.3cm}\\
%
Institut f\"ur Theoretische Physik und Astrophysik, 
Universit\"at W\"urzburg \\
Am Hubland, D-97074 W\"urzburg, Germany
%
}

\maketitle

\begin{abstract}
We discuss the recent derivation of the two loop virtual QCD
corrections to the W boson pair production in the 
quark-anti-quark-annihilation channel in the limit where all 
kinematical invariants are large compared to the mass of the W boson.
In particular, we describe the use of the {\tt PSLQ} 
algorithm on an example integral.
\end{abstract}

\section{Introduction}
\label{sec:intro}

The Large Hadron Collider (LHC) will be
the centre of interest for particle physics phenomenology in the next years.
Open issues that require definite answers are the 
verification of the consistency and
validity of the Standard Model (SM) in the energy range of the LHC
as well as insights into New Physics. 
Probably, the most important goal for the
LHC is the discovery of the elusive Higgs boson.  
Another important endeavour at the LHC 
is the precise measurement of the
hadronic production of gauge boson pairs.
Deviations from the SM predictions would indicate the presence of either
anomalous couplings or new heavy particles which would decay into
vector boson pairs~\cite{tevatron1, tevatron2}.

In this context,
W pair production via quark-anti-quark-annihilation,
\begin{equation}
\label{eq:qqWW}
q {\bar q} \rightarrow W^+ \, W^- \, ,
\end{equation}
is a very important process at the LHC.
Firstly, it will serve as a signal process 
in the search for New Physics since it can be used
to measure the vector boson trilinear couplings as predicted by the
Standard Model (SM).
Secondly,  $ q {\bar q} \rightarrow W^+ W^-$ is the dominant irreducible
background to the promising Higgs discovery channel 
$p p \rightarrow H \rightarrow W^* W^* 
\rightarrow l {\bar \nu} {\bar l}' \nu'$,
in the mass range M$_{\mathrm{Higgs}}$ 
between 140 and 180 GeV~\cite{dittmardreiner}.

Due to its importance, the study of W pair production in hadronic 
collisions has attracted 
a lot of attention in the literature~\cite{brown,ohn,fri,
dixon1,dixon2,campbell,Grazzini:2005vw} and is currently known at NLO.
Nonetheless, if a theoretical estimate for  the
W pair production is to be compared against 
experimental measurements at the LHC, one is bound to 
go one order higher in the perturbative expansion, namely
to the next-to-next-to-leading order (NNLO). This would 
allow, in principle, an accuracy at the level of 10\%.
High accuracy for the W pair production is also needed
when the process is studied as background to Higgs discovery channel
$g g \rightarrow H$.
The signal in this case is currently known at NNLO 
level~\cite{Spira:1995rr,Dawson:1990zj,Harlander:2002wh,Anastasiou:2002yz,
Ravindran:2003um,
Catani:2001cr,Davatz:2004zg,Anastasiou:2004xq,Anastasiou:2007mz,
Grazzini:2008tf,Bredenstein:2006rh}.
Another process that needs to be included in the background
is the W pair production
in the loop induced gluon fusion channel, 
$g g \rightarrow W^+ W^-$
which
contributes at $\mathcal{O}(\alpha_s^2)$ relative to the 
quark-anti-quark-annihilation channel but is 
nevertheless enhanced due to the large gluon flux
at the LHC~\cite{kauer1,kauer2}.

Here we discuss some technical details occurring
in the
computation of the NNLO virtual corrections for W pair 
production~\cite{Chachamis:2007cy, Chachamis:2008yb, Chachamis:2008xu}.
Our methodology for obtaining
the massive amplitude (massless fermion-boson scattering was studied 
in~\cite{Anastasiou:2002zn}) 
is very similar to the one followed 
in~\cite{Czakon:2007ej,Czakon:2007wk,Czakon:2004wm} which is, 
at its turn, an evolution
of the methods employed in~\cite{Czakon:2006pa,Actis:2007gi}.
The amplitude is reduced to an expression that only
contains a small number of integrals (master integrals)
with  the help of the Laporta algorithm~\cite{Laporta:2001dd}.
Next comes the construction, in a fully automatised way,
of the Mellin-Barnes (MB) 
representations~\cite{Smirnov:1999gc,Tausk:1999vh}
of all the master integrals by using
the {\tt MBrepresentation} package~\cite{MBrepresentation}. The
representations  are then
analytically continued in the number of space-time dimensions by means of the
{\tt MB} package~\cite{Czakon:2005rk}, thus revealing the full singularity
structure. An asymptotic 
expansion in the mass parameter is performed by
closing contours and the integrals are finally resummed,
either with the help of {\tt XSummer}~\cite{Moch:2005uc}
or the {\tt PSLQ} algorithm~\cite{pslq:1992}.
In this paper we give an example of the use of the latter.

\section{The Calculation}

\begin{figure}
\centerline{\includegraphics[width=0.5\columnwidth]
{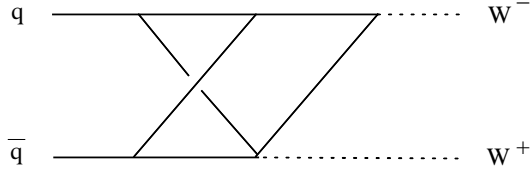}}
\caption{A non-planar two-loop master integral.}
\label{fig:master52}
\end{figure}

Consider the master integral shown in Fig.~\ref{fig:master52}.
The MB representation is
\begin{eqnarray}
\label{eq:rep}
\rm{I}_{\rm{master}} &=&
\int_{- i \infty}^{i \infty} \prod_{j = 1}^{7} d z_j
\left(-m^2\right)^{-2 \epsilon-z_1-2} (-s)^{-z_3}
(-t)^{z_1+z_2+z_3} 
(-u)^{-z_2}
 \nonumber \\ && \times
\Gamma(z_2) \Gamma (-z_1-z_2-z_3) 
\Gamma (z_3)
\Gamma (z_1+z_3+1)
\Gamma (-z_4) 
\Gamma (-z_2 +z_4+z_5+1) 
\nonumber \\ && \times
\Gamma (-2 \epsilon+z_2+z_3-z_4-z_5-1) 
\Gamma (-z_5) 
\Gamma (-2   \epsilon-z_1-z_2-z_3-z_4-z_6-1) 
\nonumber \\ && \times
\Gamma (-z_6) 
\Gamma (z_1+z_2+z_4+z_6+1) 
\Gamma (-z_3+z_5+z_6+1) 
\Gamma (-2 \epsilon+z_4-z_7)
\nonumber \\ && \times
\Gamma(\epsilon+z_1+z_4+z_5+z_6+2) 
\Gamma (-3 \epsilon-z_1-z_4-z_5-z_6-z_7-2) 
\Gamma (-z_7) 
\nonumber \\ && \times
\Gamma (-\epsilon-z_4+z_7) 
\Gamma (2 \epsilon+z_1+z_4+z_5+z_6+z_7+2)
\nonumber \\ && \times
\left(
\Gamma (-3 \epsilon) 
\Gamma (-4 \epsilon-z_1-2 z_4-z_5-z_6-2) 
\Gamma (z_1+z_5+z_6+2)
\right. \nonumber \\ && \times \left. 
\Gamma (z_1+2 z_4+z_5+z_6+2) 
\Gamma (-2 \epsilon-z_7)
\right)^{-1}\,,
\end{eqnarray}
where, s, t, u the Mandelstam variables and m the mass of the W.

One of the most important steps in our calculation is the computation
of the master integrals in closed analytic form after expansion in the 
mass.
For that, one needs
in most cases to resum integrals that are constants and which appear
after reducing the dimensionality of the initial MB representations
by applying Barnes' Lemmas. The constants are expressed in terms
of the transcendental basis
$\left \{ 1,\, \zeta_2 = \pi^2/6,\, \zeta_3,\, \zeta_4 \right \}$.

While performing the steps described in the last paragraph of the
Introduction for the computation of $\rm{I}_{\rm{master}}$, 
one needs to resum the following integral
\begin{eqnarray}
\label{eq:integral}
\rm{I}_{\rm{const}} &=&
\int_{c_1 - i\infty }^{c_1 + i\infty } dz_1 
\int_{c_2 - i\infty }^{c_2 + i\infty } dz_2 
\frac{1}{4 (z_2-1) z_2}
\left(\Gamma (1-z_1) \Gamma (z_1) \Gamma (1-z_2) 
\Gamma (-z_1-z_2+1) 
\right. \nonumber \\ && \left.
\Gamma (z_2)  \Gamma (z_1+z_2) \left(-\psi ^{(0)}(-z_1)-4 (z_1-z_2) \psi
^{(0)}(-z_1-z_2+1)
\right. \right. \nonumber \\ && \left. \left.
+4 z_1 \psi ^{(0)}(-z_1-z_2+2)
-4 z_2 \psi ^{(0)}(-z_1-z_2+2)+\psi ^{(0)}(z_2-1)\right)\right),
\end{eqnarray}
where $\psi ^{(0)}(x)$ is the digamma function, 
$c_1 = -1/3$, $c_2 = -1/3$ and
the integration contours are straight lines.

Using a version of the double exponential algorithm~\cite{mori},
it takes about five minutes to integrate
numerically Eq.~\ref{eq:integral} and get
to about 30 digits.
\begin{eqnarray}
\label{eq:num}
\rm{I}_{\rm{const}}^{\rm{num}} = 0.6265076409145894496726188621488\,.
\end{eqnarray}
Next, using the {\tt PSLQ} algorithm  
we fit $\rm{I}_{\rm{const}}^{\rm{num}}$ in
terms of our basis and we finally obtain
\begin{eqnarray}
\rm{I}_{\rm{const}} = -\frac{1}{2}+\frac{7 \pi ^2}{24}-\frac{5
\zeta (3)}{2}\,.
\end{eqnarray}
The numerical value of the resummed $\rm{I}_{\rm{const}}$ calculated 
with 32 significant digits accuracy is
\begin{eqnarray}
\rm{N}[\rm{I}_{\rm{const}}, 32] = -0.62650764091458944967261886214808
\end{eqnarray}
and
one can see that it
agrees with the value in Eq.~\ref{eq:num} for the first 30 digits.

\section{Conclusions and Outlook}
We have discussed some details of the computation
of the two-loop NNLO QCD virtual corrections for
the process $q {\bar q} \rightarrow W^+ \, W^-$ in the limit of small vector
boson mass. 
Our main result was presented in~\cite{Chachamis:2008yb}.
This was a first step towards the
complete evaluation of the
virtual corrections. In a forthcoming publication, we will derive a series
expansion in the mass and integrate the result numerically to recover the
full mass dependence, similarly to what has been done in~\cite{Czakon:2008zk}.

To complete the NNLO project one still needs to consider 
$2 \to 3$ real-virtual contributions  
and $2 \to 4$ real ones. The real-virtual corrections
are known from the NLO studies on $W W + jet$ production
in~\cite{Campbell:2007ev,
Dittmaier:2007th}.

\section{Acknowledgments}

This work was supported by the Sofja Kovalevskaja Award of the
Alexander von Humboldt Foundation.


\begin{footnotesize}



%

\end{footnotesize}


\end{document}